\documentclass[12pt]{iopart}

\usepackage{iopams}
\usepackage{graphicx}

\begin{document}

\title[Stark effect in the nonuniform field]%
{Stark effect in the nonuniform field and its influence on
the fine structure of Rydberg blockade}

\author{Yurii V Dumin$^{1,2,3}$}

\address{$^1$ Max Planck Institute
for the Physics of Complex Systems (MPIPKS),\\
Noethnitzer Str.\ 38, 01187 Dresden, Germany}

\address{$^2$ Sternberg Astronomical Institute (GAISh)
of Lomonosov Moscow State University,\\
Universitetski prosp.\ 13, 119992 Moscow, Russia}

\address{$^3$ Space Research Institute (IKI)
of Russian Academy of Sciences,\\
Profsoyuznaya str.\ 84/32, 117997 Moscow, Russia}

\eads{\mailto{dumin@pks.mpg.de}, \mailto{dumin@yahoo.com}}

\begin{abstract}
Splitting the energy levels of a hydrogen-like atom by the electric
field nonuniform at the atomic scale is studied.
This situation is important for the multi-level treatment of
the phenomenon of Rydberg blockade [Yu.V. Dumin, \textit{J.\ Phys.\ B},
v.\,47, p.\,175502 (2014)].
An explicit formula for the energy levels is derived.
A typical value of the energy shift by the electric field gradient turns
out to be proportional to the 4th power of the principal quantum number
(\textit{i.e.}, the square of atomic size), as would be expected
from a qualitative consideration.
Finally, the fine spatial structure of the Rydberg blockade is analyzed
when the electric-field-gradient term plays the dominant role,
and the results are confronted with the experimental data.
\end{abstract}


\submitto{\jpb}

\maketitle

\section{Introduction}
\label{sec:Introduction}

Splitting the atomic energy levels by an external electric field
(Stark effect) is studied both experimentally and theoretically for almost
a century (\textit{e.g.}, reviews~\cite{Bet_Sal,Gallagher}).
However, almost all previous theoretical treatments were performed in
the approximation of the uniform external field~$ {\cal E}_0 $.
This is not surprising because, until recently, it was difficult to imagine
that the electric field can be variable at the atomic scale.

The only work beyond this approximation, familiar to us, was done in 1970
by Bekenstein and Krieger~\cite{Bekenstein70}, who assumed that
the substantially nonuniform electric fields could be produced just by
the neighboring atoms in a sufficiently dense gas.
So, the corresponding Stark effect would result in the specific broadening
of the spectral lines.
Unfortunately, deriving the parameters of the nonuniform Stark effect
from the line broadening is a quite indirect and ambiguous procedure;
so that this idea did not attract a sufficient attention at that time.

However, the situation changed in the recent decade in the context of
the so-called Rydberg blockade, which was discussed, at first,
theoretically~\cite{Lukin01} and then confirmed
experimentally~\cite{Tong04,Singer04,Weidemueller09}.
Briefly speaking, this effect occurs when the electric field of
an initially-excited Rydberg atom disturbs the energy levels of
the neighboring ground-state atoms, so that their excitation by
a narrow-band laser irradiation to the same quantum state becomes
impossible (because this state goes out of the resonance); a pictorial
illustration can be found below in figure~\ref{fig:Rydberg_blockade}.

Although the phenomenon of Rydberg blockade is usually considered in
the approximation of the selected essential states, a reasonable alternative
treatment can be based on the consideration of Stark effect produced by
an already excited Rydberg atom on the surrounding atoms.
A particular advantage of such an approach is that it can reveal
complex spatial structure of the blockade zone, namely, a sequence of
intermittent co-centric shells where the possibility of excitation
becomes blocked and unblocked again~\cite{Dumin14}.

However, usage of the standard formulas for Stark effect seems to be
questionable in the above-mentioned situation: the characteristic
interatomic separation is only a few times greater than a typical size
of the Rydberg atom and, therefore, the electric field becomes substantially
nonuniform at the atomic scale.
So, it is desirable to perform the corresponding analysis taking into
account Stark splitting by the nonuniform external field.

Below, in section~\ref{sec:Stark_Effect}, we first briefly remind
the basic mathematical formalism for dealing with Stark effect by
the quantization in parabolic coordinates and then apply the perturbation
technique to the case of the nonuniform electric field.
(As distinct from the above-cited paper~\cite{Bekenstein70}, our approach
will not use the WKB approximation.)
At last, in section~\ref{sec:Rydberg_blockade}, we study the Rydberg
blockade in the situation when it is produced mostly by the gradient term
of the Stark splitting and discuss the respective observational
consequences.

\section{Stark effect in the nonuniform external field}
\label{sec:Stark_Effect}

\subsection{Basic formulas for the unperturbed Coulomb's problem}
\label{subsec:Basic_formulas}

The starting point of our consideration is Schroedinger equation
for an electron in the Coulomb's field of the nucleus with charge~$ Z $,
located in the origin of coordinates, and the external electric
field~$ \cal E $, directed along $ z $-axis
(all formulas are written in the atomic units):
\begin{equation}
\Big[ \! - \! \frac{1}{2} \, \triangle \, - \frac{Z}{r} \, + \, \delta U (z)
  \Big] \psi = E \psi \: ,
\label{eq:Schroedinger}
\end{equation}
where $ \delta U $ is perturbation of the atomic potential energy by
the electric field
\begin{equation}
{\cal E} (z) = \, {\cal E}_0 +
  \Bigg( \! \frac{d {\cal E}}{dz} \! {\Bigg)}_{\!\! 0} z \, + \, \dots \: ,
\label{eq:Electric_field}
\end{equation}
where subscript 0 denotes the corresponding values in the origin of
coordinates.
So, the perturbation in explicit form is
\begin{equation}
\delta U (z) = \, {\cal E}_0 z + \frac{1}{2} \,
  \Bigg( \! \frac{d {\cal E}}{dz} \! {\Bigg)}_{\!\! 0} z^2 + \, \dots
\label{eq:Potential_perturb}
\end{equation}
(It is written with a plus sign because it refers to the negatively-charged
electron.)

If $ \delta U \! \equiv 0 $, then equation~(\ref{eq:Schroedinger})
can be solved exactly in the parabolic coordinates in terms of
either the confluent hypergeometric functions (\textit{e.g.},
textbook~\cite{Lan_Lif_v3}, {\S}37) or the associated Laguerre functions
~$ L_{\lambda}^{\mu} $ (monograph~\cite{Bet_Sal}, section~6).
We prefer to use the second option, so the solution will take
the form:
\begin{eqnarray}
&& {\psi}_{n_1 n_2 m}^{(0)} ( \xi , \eta , \varphi ) =
  \frac{ e^{ \pm i m \varphi } }{ \sqrt{ \pi n } } \,
\nonumber \\
&& \qquad \times \!
  \frac{ ({n_1}!)^{1/2} }{ ((n_1 + m)!)^{3/2} } \,
  \frac{ ({n_2}!)^{1/2} }{ ((n_2 + m)!)^{3/2} } \:
  {\varepsilon}^{ m + \frac{3}{2} } \,
\nonumber \\[1ex]
&& \qquad \times \!
  e^{ - \varepsilon ( \xi + \eta ) / 2 }
  ( \xi \eta )^{m/2}
  L^{m}_{ n_1 + m } ( \varepsilon \xi ) \,
  L^{m}_{ n_2 + m } ( \varepsilon \eta ) \, .
\label{eq:Wave_function_exact}
\end{eqnarray}
Here, $ \varepsilon = Z / n $,
$ \varphi $~is the azimuthal angle; and
$ \xi , \eta $~are the parabolic coordinates, related to
the Cartesian coordinates by the standard formulas:
\numparts
\begin{eqnarray}
\xi     & = & r + z \: ,
\label{eq:Parab_coord_xi} \\
\eta    & = & r - z \: ,
\label{eq:Parab_coord_eta} \\
\varphi & = & \arctan ( x / y ) \: ,
\label{eq:Parab_coord_phi}
\end{eqnarray}
\endnumparts
\begin{equation}
r = \sqrt{ x^2 + y^2 + z^2 } = \frac{1}{2} \, ( \xi + \eta ) \: ;
\label{eq:Parab_coord_r}
\end{equation}
or \textit{vice versa}:
\numparts
\begin{eqnarray}
x & = & \sqrt{ \xi \eta } \, \cos \varphi \: ,
\label{eq:Parab_coord_x} \\
y & = & \sqrt{ \xi \eta } \, \sin \varphi \: ,
\label{eq:Parab_coord_y} \\
z & = & \frac{1}{2} \, ( \xi - \eta ) \: ,
\label{eq:Parab_coord_z}
\end{eqnarray}
\endnumparts
where
\begin{equation}
\xi , \eta \, \in [ \, 0 , +\infty ) \, , \quad
\varphi \, \in [ \, 0 , 2 \pi ] \: .
\label{eq:Range_eta_phi}
\end{equation}
The elementary length and volume in the parabolic coordinates are
\begin{eqnarray}
dl^2 \! & = & \frac{ \xi + \eta }{ 4 \xi } \, d{\xi}^2 \, + \,
  \frac{ \xi + \eta }{ 4 \eta } \, d{\eta}^2 \, + \:
  \xi \eta \, d{\varphi}^2 ,
\label{eq:Parab_coord_Length} \\[0.5ex]
dV \! & = & \frac{ \xi + \eta }{4} \, d{\xi} \, d{\eta} \, d{\varphi} \, .
\label{eq:Parab_coord_Volume}
\end{eqnarray}

The eigenvalues~$ E^{(0)}$ of the unperturbed equation~(\ref{eq:Schroedinger})
are associated with the principal quantum number~$ n $ by the usual
formula:
\begin{equation}
n = \frac{1}{ \sqrt{ - 2 E^{(0)}} } \, , \quad
n = 1, 2, 3, \dots \: ;
\label{eq:Princ_quant_numder_Energy}
\end{equation}
while the principal quantum number is related to the parabolic quantum
numbers~$ n_1 $ and $ n_2 $ (which are the nonnegative integers) and
the magnetic quantum number~$ m $ as
\begin{equation}
n = n_1 + n_2 + m + 1 \, ,
\label{eq:Relation_Quant_numders}
\end{equation}
where $ m $ is assumed to be positive or zero, since the plus/minus sign
was already written explicitly in~(\ref{eq:Wave_function_exact}).

\subsection{Perturbation by the external field}
\label{subsec:Nonuniform_field}

Let us find eigenvalues of the Schroedinger equation~(\ref{eq:Schroedinger})
perturbed by the nonuniform external electric field~(\ref{eq:Electric_field}).
As was already mentioned in the Introduction, a similar problem was
treated long time ago by Bekenstein and Krieger~\cite{Bekenstein70}, who
developed a quite sophisticated mathematical technique based on
the WKB approximation.
So, our aim here is to perform the same calculations beyond
the above-mentioned approximation and, thereby, to check and refine
the earlier results.

If $ ( d {\cal E} \! / dz \! )_{0} \! = \! 0 $, \textit{i.e.}, the external
electric field is uniform, then the first-order correction to the energy
levels is well known~\cite{Bet_Sal,Gallagher,Lan_Lif_v3}:
\begin{equation}
( \delta E )_{\rm unif} = \frac{3}{2} \: {\cal E} n \, ( n_1 - n_2 ) \, .
\label{eq:Delta_E_0}
\end{equation}

In the more general case, when $ ( d {\cal E} \! / dz \! )_{0} \neq 0 $,
\textit{i.e.}, the electric field gradient is present, the perturbation of
the energy levels~$ \delta E $ is given in the first approximation by
the diagonal matrix element of operator~(\ref{eq:Potential_perturb})
with respect to the unperturbed states~(\ref{eq:Wave_function_exact}):
\begin{equation}
\delta E  = \! \int \!\!\! \int \!\!\! \int \! | {\psi}_{ n_1 n_2 m } |^2
  ( \delta U ) \, dV
\label{eq:Delta_E_gen}
\end{equation}
(for rigorous mathematical justification of using the first-order
perturbation theory in the nonuniform Stark problem, see~\cite{Bekenstein70}
and references therein).

Substituting expression~(\ref{eq:Potential_perturb})
into~(\ref{eq:Delta_E_gen}) and using the properties of parabolic
coordinates~(\ref{eq:Parab_coord_z}) and~(\ref{eq:Parab_coord_Volume}),
we get:

\begin{eqnarray}
&& \delta E = \frac{1}{8} \int\limits_0^{\: \infty} \!\!
  \int\limits_0^{\: \infty} \!\! \int\limits_0^{\; \; 2 \pi}
  | {\psi}_{ n_1 n_2 m } |^2 \,
  \bigg[ \, {\cal E}_0 ( {\xi}^2 \! - {\eta}^2 )
\nonumber \\[-0.5ex]
&& \qquad \qquad \qquad
  + \frac{1}{4} \, \Bigg( \! \frac{d {\cal E}}{dz} \!
  {\Bigg)}_{\!\! 0} ( {\xi}^2 \! - {\eta}^2 ) ( \xi - \eta ) \bigg]
  d \varphi \, d \xi \, d \eta \, .
\label{eq:Delta_E_modif1}
\end{eqnarray}

Then, the total energy of the split sublevels can be written as
\begin{equation}
E_{n_1 n_2 m} = \, - \frac{1}{2} \, \frac{Z^{\, 2}}{n^2} \, + \,
  (\delta E)_{ n_1 n_2 m } \, ,
\label{eq:Energy_total}
\end{equation}
where
\begin{eqnarray}
&& (\delta E)_{ n_1 n_2 m } = \frac{1}{4n} \, \frac{ n_1 ! }{ ( n_1 + m)! } \,
  \frac{ n_2 ! }{ ( n_2 + m)! } \,
\nonumber \\[1ex]
&& \quad \times \!\! \Bigg\{ \frac{ {\cal E}_0 }{ \varepsilon } \,
  \Big[ J_{n_1 + m, \, m}^{(2)} J_{n_2 + m, \, m}^{(0)} -
  J_{n_1 + m, \, m}^{(0)} J_{n_2 + m, \, m}^{(2)} \Big]
\nonumber \\[1ex]
&& \quad + \frac{ ( d {\cal E} \! / dz \! )_{0} }{ 4 \, {\varepsilon}^2 } \,
  \Big[ J_{n_1 + m, \, m}^{(3)} J_{n_2 + m, \, m}^{(0)}
  + J_{n_1 + m, \, m}^{(0)} J_{n_2 + m, \, m}^{(3)}
\nonumber \\[1ex]
&& \quad - J_{n_1 + m, \, m}^{(2)} J_{n_2 + m, \, m}^{(1)}
  - J_{n_1 + m, \, m}^{(1)} J_{n_2 + m, \, m}^{(2)} \Big] \Bigg\} \, .
\label{eq:Energy_perturb}
\end{eqnarray}
Here, the integrals~$ J_{ \lambda \mu }^{ ( \sigma ) } $ are defined
by the standard way as
\begin{equation}
J_{ \lambda \mu }^{ ( \sigma ) } = \frac{1}{ ({\lambda}!)^2 }
  \int\limits_0^{ \, \infty } e^{-{\rho}} {\rho}^{ \, {\mu} + {\sigma} } \,
  \big[ L_{\lambda}^{\mu} ({\rho}) {\big]}^2 \, d {\rho} \, .
\label{eq:Int_J_def}
\end{equation}

Their explicit calculation gives the following expression in terms of
the binomial coefficients (\textit{e.g.}, \cite{Bet_Sal}, section~3):
\begin{eqnarray}
J_{ \lambda \mu }^{ ( \sigma ) } \! & = &
  (-1)^{\sigma} \frac{ {\lambda}! }{ ( {\lambda} - {\mu} )! } \: {\sigma}!
\nonumber \\
& \times & \!\! \sum\limits_{ {\beta} = 0 }^{\sigma} (-1)^{\beta}
  \Big( \! \begin{array}{c} \sigma \\[-0.5ex] \beta \end{array} \! \Big)
  \Big( \!\! \begin{array}{c} \lambda \! + \! \beta \\[-0.5ex] \sigma
  \end{array} \! \Big)
  \Big( \!\! \begin{array}{c} \lambda \! + \! \beta \! - \! \mu \\[-0.5ex]
  \sigma \end{array} \! \Big) \, .
\label{eq:Int_J_explicit}
\end{eqnarray}
In particular, the integrals required for us are
\numparts
\begin{eqnarray}
J_{k+m, \, m}^{(0)} \! & = & \frac{(k \! + \! m)!}{k!} \, ,
\label{eq:J_0}
\\[1ex]
J_{k+m, \, m}^{(1)} \! & = & \frac{(k \! + \! m)!}{k!} \,
  ( 2k + m + 1 ) ,
\label{eq:J_1}
\\[1ex]
J_{k+m, \, m}^{(2)} \! & = & \frac{(k \! + \! m)!}{k!} \,
  \big[ ( k + m )^2 + ( k + m ) ( 4k + 3 )
  + ( k^2 \! + 3k + 2 ) \big] \, ,
\label{eq:J_2}
\\[1ex]
J_{k+m, \, m}^{(3)} \! & = & \frac{(k \! + \! m)!}{k!} \,
  \big( 20 k^3 \! + 30 k^2 \! + 22 k + 30 k^2 m + 12 k m^2 \! + 30 k m
\nonumber \\[0.5ex]
  && \qquad \quad \; + m^3 \! + 6 m^2 \! + 11 m + 6 \big) \, .
\label{eq:J_3}
\end{eqnarray}
\endnumparts

After the substitution of~(\ref{eq:J_0})--(\ref{eq:J_3})
into~(\ref{eq:Energy_perturb}) and~(\ref{eq:Energy_total}),
the final result for splitting the energy levels can be written
in a quite compact form as
\begin{eqnarray}
&& E_{n_1 n_2 m} = - \frac{1}{2} \, \frac{Z^{\, 2}}{n^2} \, + \,
   \frac{3}{2} \: {\cal E}_0 \frac{n}{Z} \, ( n_1 \! - \! n_2 )
\nonumber \\[1ex]
&& \quad
  + \frac{1}{4} \, \Bigg( \! \frac{d {\cal E}}{dz} \! {\Bigg)}_{\!\! 0} \,
  \frac{n^2}{Z^{\, 2}} \, \Big[
    5 ( n_1 \! - \! n_2 )^2 +
    2 ( n_1 n_2 + 1 ) +
    ( n \! - \! m ) \, m +
    n_1 \! +
    n_2
  \Big] ,
\label{eq:Final_result}
\end{eqnarray}
where $ n = n_1 \! + n_2 \! + m + 1 $
($ n_1 \! \geq 0 , \: n_2 \! \geq 0 , \: m \! \geq 0 $).
Here, the first term represents the energy of an unperturbed hydrogen-like
atom, the second term is the well-known expression for linear Stark effect
in the uniform external field~\cite{Bet_Sal,Gallagher}, and the third
(gradient) term is the required correction for nonuniformity.
To avoid misunderstanding, let us emphasize that this gradient term
should not be mixed with the higher-order corrections with respect to
the electric field amplitude~$ {\cal E}_0 $, which were widely discussed
in the previous literature (\textit{e.g.}, the procedure for obtaining
the corrections of an arbitrary order can be found
in~\cite{Silverstone78}).

\subsection{Qualitative features}
\label{subsec:Qualitative_features}

Let us briefly discuss a qualitative behavior of the gradient term.
First of all, it should be noticed that the electric field gradient
partially lifts a degeneracy of the energy levels with respect to
the magnetic quantum number~$ m $
(apart from the plus/minus sign in formula~(\ref{eq:Wave_function_exact}))
already in the first order of the perturbation theory, while in the uniform
field this is possible only in the second order~\cite{Bet_Sal,Gallagher}.

Next, as is known, the linear Stark effect in a uniform field is roughly
proportional to~$ n^2 {\cal E}_0 $, which is just a typical difference of
the electric potential across the atom.
Besides, this quantity can be substantially reduced under appropriate
choice of the quantum numbers, \textit{e.g.}, when the atom is approximately
symmetric ($ n_1 \approx n_2 $).

The gradient term exhibits a quite similar behavior: as follows
from~(\ref{eq:Final_result}), its characteristic magnitude
is~$ ( d {\cal E} \! / dz \! )_{0} \, n^4 $, which represents again
the typical potential difference across the atom.
This value can also be substantially reduced at the appropriate
combination of the quantum numbers.
However, the sum in the square brackets always remains positive, so that
the sign of the resulting effect is completely determined by the sign of
the derivative $ ( d {\cal E} \! / dz \! )_{0} $.

At last, let us make a few remarks about the application of our results
to other Rydberg atoms than hydrogen, namely, with presence of
the quantum defects.
A detailed study of this issue, evidently, requires a separate paper.
However, it can be reasonably believed that difference in the gradient
term of the Stark effect between the purely hydrogenic and hydrogen-like
atoms will not be so dramatic as in the case of the uniform field.
Really, the crucial feature of Stark effect in the hydrogen-like atoms
with non-degenerate spectra is that it is quadratic with respect to
the field amplitude~$ {\cal E}_0 $:
the first-order correction to energy
$ {\cal E}_0 < \!\! \psi | \, z \, | \psi \!\! > $
vanishes because the wavefunction~$ | \psi \!\! > $ possesses
a definite parity under the transformation $ z \to -z $,
while the operator~$ z $ changes its sign.
So, the matrix element
$ < \!\! \psi | \, z \, | \psi \!\! > $
can be nonzero only for
pure hydrogen, where the energy spectrum is degenerate and, therefore,
$ | \psi \!\! > $~can be a superposition of states with various parities.

On the other hand, when we calculate the gradient correction,
which is proportional to~$ z^2 $, the difference should not be so drastic:
since operator~$ z^2 $ does not change its sign under the transformation
$ z \to -z $, the first-order correction
$ (1/2) ( d {\cal E} \! / dz \! )_{0} < \!\! \psi | \, z^2 \, | \psi \!\! > $,
in general, will not be equal to zero even in the non-degenerate state with
a certain parity.
Therefore, there should be not so much difference between the purely
hydrogenic and hydrogen-like atoms in the case of the nonuniform-field
Stark effect.

\subsection{Comparison with the earlier calculations}
\label{subsec:Comparison}

A convenient benchmark for testing our calculations are the results by
Bekenstein and Krieger (BK)~\cite{Bekenstein70}, which were derived in the WKB
approximation.
They can be rewritten in our designations as:
\begin{eqnarray}
&& \delta E^{\rm (BK)} {\Big|}_{{\cal E}_0 = 0} =
  \frac{1}{4} \,
  \Bigg( \!\! \frac{\partial {\cal E}_z}{\partial z} \!\! \Bigg)_{\!\!\! 0}
  \frac{n^2}{Z^2} \,
\nonumber \\[1ex]
&& \quad \times \!\! \Big[
    5 ( n_1 \! - \! n_2 )^2 -
    2 ( \, 2 n_1 n_2 \! + n_1 \! + n_2 )
\nonumber \\
&& \quad + m ( \, m \! - \! 2 n ) \Big]
\nonumber \\[1ex]
&& \quad + \frac{3}{4} \,
  \Bigg[
  \Bigg( \!\! \frac{\partial {\cal E}_x}{\partial x} \!\! \Bigg)_{\!\!\! 0} \!
  - \Bigg( \!\! \frac{\partial {\cal E}_y}{\partial y} \!\! \Bigg)_{\!\!\! 0}
  \Bigg]
  {\epsilon}_m \, \frac{n^2}{Z^2} ( n_1 \! + 1 ) ( n_2 \! + 1 ) \, ,
\label{eq:Bekenstein}
\end{eqnarray}
where
\begin{equation}
{\epsilon}_m =
  \left\{ \!\!
    \begin{array}{rcl}
      \pm 1 & \mbox{at} & \! m = 1 ,    \\
      0     & \mbox{at} & \! m \neq 1 ,
    \end{array}
  \right.
\label{eq:epsilon_def}
\end{equation}
and the plus/minus sign corresponds to the sign of~$ m $ in
the wavefunction~(\ref{eq:Wave_function_exact}).

The second term in formula~(\ref{eq:Bekenstein}) vanishes for
any axially-symmetric field, when
$ \partial {\cal E}_x / \partial x = \partial {\cal E}_y / \partial y $.
In particular, the respective components of the electric-field-gradient
tensor were ignored in our initial formulation of the problem.
So, only the first term of~(\ref{eq:Bekenstein}) should be confronted
with our formula~(\ref{eq:Final_result}).

It is seen that these two expressions, in general, are somewhat different
from each other.
However, the first term in the square brackets is the same.
So, in the case most important from the experimental point of view,
when the magnetic quantum number~$ m $ is small and the atomic state
is strongly asymmetric (\textit{i.e.}, one of the parabolic quantum
numbers, $ n_1 $ or $ n_2 $, is close to zero and another is approximately
equal to the principal quantum number~$ n \gg 1 $), both formulas give
the same result:
$ (5/4) (d { \cal E} \! / \! dz )_0 ( n^4 \! / Z^2 ) $.

On the other hand, the WKB approximation~(\ref{eq:Bekenstein}) substantially
deviates from the exact solution~(\ref{eq:Final_result}) for the symmetric
states, where $ n_1 \! \approx \! n_2 $.
Moreover, this approximation leads to the conclusion that
the gradient-term Stark effect can change its sign in such states,
which is very suspicious from the physical point of view.
Besides, as follows from these formulas, dependence on the magnetic
quantum number~$ m $ in the WKB approximation is inadequate.

So, the analysis performed in section~\ref{subsec:Nonuniform_field}
enabled us to improve substantially the previously-known formula
for Stark effect in the nonuniform field.
As will be seen below in section~\ref{subsec:Gradient_term},
the established properties (in particular, the impossibility to
change the sign of the total effect) lead to important qualitative
features in the pattern of Rydberg blockade.

\section{Rydberg blockade}
\label{sec:Rydberg_blockade}

As was already mentioned in the Introduction, the most interesting
application of Stark effect in the nonuniform field is
a multi-level treatment of the Rydberg blockade, which was performed
for the case of a uniform field in our previous article~\cite{Dumin14}.
The same analysis taking into account all terms in
formula~(\ref{eq:Final_result}) is quite straightforward but cumbersome
and, therefore, requires a separate paper.
So, we shall restrict our consideration here by the case when the gradient
term plays a dominant role.
From the physical point of view, this situation assumes a strong
electric-field gradient $ ( d {\cal E} \! / dz \! )_{0} $ and/or very
large values of the principal quantum number~$ n $.

\subsection{Blockade by the gradient term}
\label{subsec:Gradient_term}

We shall consider below a neutral hydrogen-like atom, \textit{i.e.},
take $ Z = 1 $.
Besides, since the experiments on Rydberg blockade are typically
performed with atoms possessing small values of the magnetic quantum
number ($ m = 0, 1, 2 $), it is reasonable to neglect the terms
with~$ m $ in formula~(\ref{eq:Final_result}).
Within the same accuracy, we can discard also the terms on the order of
unity as compared to~$ n $.
At last, expressing~$ n_2 $ in terms of $ n $ and $ n_1 $,
the energy of the split sublevels is written as
\begin{equation}
E_{n n_1} \approx - \frac{1}{2} \, \frac{1}{n^2} \,
  + \, \frac{1}{4} \, \Bigg( \! \frac{d {\cal E}}{dz} \! {\Bigg)}_{\!\! 0}
  n^2 \Big[ 5 n^2 - 18 n_1 \, ( n - n_1 ) \Big] .
\label{eq:Energy_simpl}
\end{equation}

Next, it can be easily shown that the expression in square brackets
is always positive and, as function of~$ n_1 $, takes a maximum value
at the boundaries of the domain of definition, for example,
at $ n_1 = 0 $ (\textit{i.e.}, in the case of the most asymmetric atom).
So, the energy of \textit{the most perturbed sublevel} with
principal quantum number~$ n $ takes the form:
\begin{equation}
E_n^{\rm (max)} = - \frac{1}{2} \, \frac{1}{n^2} \, + \,
  \frac{5}{4} \, \Bigg( \! \frac{d {\cal E}}{dz} \! {\Bigg)}_{\!\! 0} n^4 \, .
\label{eq:Energy_most_perturb}
\end{equation}
Following the same procedure as in our previous paper~\cite{Dumin14},
just this sublevel will be used to estimate the characteristic parameters
of the Rydberg blockade zone.

Similarly to the above-cited article, we assume that the dipolar electric
field produced by an already excited Rydberg atom, located in the origin
of coordinates, is
\begin{equation}
{\cal E}(r) =
  \frac{ C n^2 }{ r^3 } \, ,
\label{eq:El_field}
\end{equation}
where $ C $~is a dimensionless coefficient on the order of unity,
which can be both positive and negative depending on the dipole orientation.
(Since the atomic system of units is used everywhere in the present work,
we shall not write tildes as in the previous paper~\cite{Dumin14}.)

Without going into details of the angular dependence, the derivative of
the electric field can be roughly estimated as
\begin{equation}
\Bigg( \! \frac{d {\cal E}}{dz} \! {\Bigg)}_{\!\! 0} \approx
  \pm \Bigg( \! \frac{d {\cal E}}{dr} \! {\Bigg)}_{\!\! 0} =
  \mp \frac{ 3 \, C n^2 }{ r^4 } \, .
\label{eq:Derivative}
\end{equation}
Here, the upper and lower signs correspond to the cases when
the perturbed atom is located, respectively, in the positive and
negative direction of $z$-axis; and subscript~0 refers, as in
section~\ref{sec:Stark_Effect}, to the position of the atom
experiencing the Stark effect.

To be specific, let us assume that the value of the derivative
$ ( d {\cal E} \! / dz \! )_{0} $ is positive:
\begin{equation}
\Bigg( \! \frac{d {\cal E}}{dz} \! {\Bigg)}_{\!\! 0} =
  \frac{ 3 \, |C| \, n^2 }{ r^4 } \, ;
\label{eq:Deriv_expl}
\end{equation}
so that expression~(\ref{eq:Energy_most_perturb}) is reduced to
\begin{equation}
E_n^{\rm (max)} = \, - \frac{1}{2} \, \frac{1}{n^2} \, + \,
  \frac{ 15 |C| }{ 4 } \, \frac{ n^6 }{ r^4 } \, .
\label{eq:Energy_distance}
\end{equation}

\begin{figure}
\center{\includegraphics[width=9.5cm]{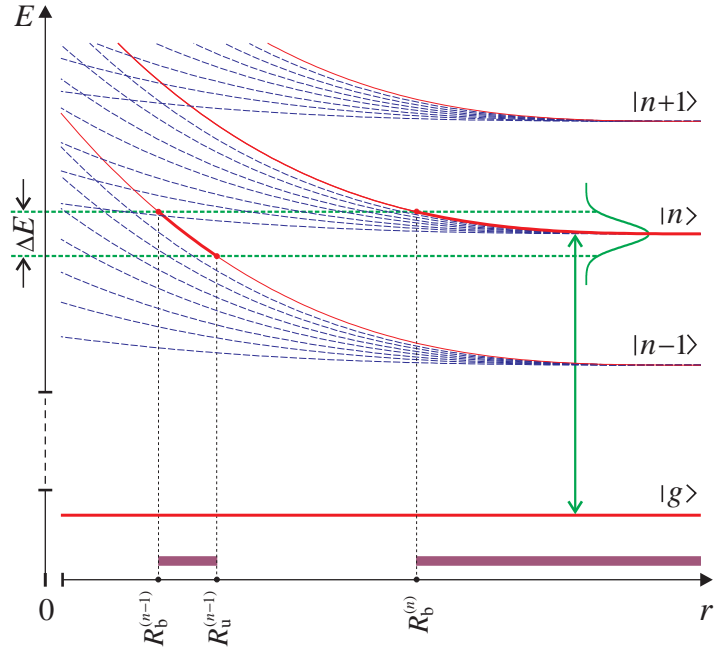}}
\caption{\label{fig:Rydberg_blockade}
The multi-level pattern of Rydberg blockade caused by
the gradient term.
The energy levels of the atom under consideration are
assumed to be split by the electric field of the central
($ r \! = \! 0 $) Rydberg-excited atom.
The sublevels with maximal splitting are shown by solid (red)
curves; and other ones, by the broken (blue) curves.
The Rydberg excitation is possible only in the thick segments
of the energy curves, located between the dotted (green) horizontal
lines, which show a characteristic bandwidth of the exciting
irradiation.
The thick (violet) strips near the horizontal axis designate
the corresponding intervals of radius:
$ [R_{\rm b}^{(n)}, +\infty ] $,
$ [R_{\rm b}^{(n-1)}, R_{\rm u}^{(n-1)}] $,
\textit{etc}.}
\end{figure}

The entire Stark manifold is schematically drawn in
figure~\ref{fig:Rydberg_blockade}.
As distinct from the pattern for the uniform field~\cite{Dumin14},
all the sublevels are shifted in the same direction
(upwards if $ ( d {\cal E} \! / dz \! )_{0} > 0 $ or
downwards if $ ( d {\cal E} \! / dz \! )_{0} < 0 $).
Therefore, in the first case, only the levels with lower values of
the principal quantum number
($ | n \! - \! 1 \rangle $, $ | n \! - \! 2 \rangle $,
\textit{etc.}) can be unblocked at the sufficiently small distances.

In the same way as in paper~\cite{Dumin14}, the condition of
Rydberg blockade of the basic level~$ | n \rangle $ at
the distance~$ R_{\rm b}^{(n)} $ can be written as
\begin{equation}
- \frac{1}{2} \, \frac{1}{n^2} \, + \,
  \frac{ 15 |C| }{ 4 } \: \frac{ n^6 }{ \big( R_{\rm b}^{(n)} \big)^4 } \, =
  - \frac{1}{2} \, \frac{1}{n^2} \, + \, \frac{1}{2} \, \Delta E \, ;
\label{eq:Blockade_n}
\end{equation}
so that
\begin{equation}
\Delta E =
  \frac{ 15 |C| }{ 2 } \: \frac{ n^6 }{ \big( R_{\rm b}^{(n)} \big)^4 } \, ,
\label{eq:Delta_E}
\end{equation}
where $ \Delta E $ is the characteristic bandwidth of the exciting
radiation.

Next, the condition of unblocking and subsequent blocking of the lower
state $ | n \! - \! 1 \rangle $ at the radii $ R_{\rm u}^{(n-1)} $ and
$ R_{\rm b}^{(n-1)} $, respectively, takes the form:
\begin{equation}
- \frac{1}{2} \, \frac{1}{ (n \! - \! 1)^2 } \, + \,
  \frac{ 15 |C| }{ 4 } \:
  \frac{ (n \! - \! 1)^6 }{ \big( R_{\rm u,b}^{(n-1)} \big)^4 }
  \, = - \frac{1}{2} \, \frac{1}{n^2} \, \mp \, \frac{1}{2} \, \Delta E \, .
\label{eq:Blockade_n-1}
\end{equation}
Substituting here expression~(\ref{eq:Delta_E}) and neglecting the terms
on the order of~$ 1/n $ as compared to unity, the above formula is reduced to
\begin{equation}
\frac{1}{ \big( R_{\rm u,b}^{(n-1)} \big)^4 } \, = \,
  \frac{4}{ 15 |C| } \: \frac{1}{n^9} \, \mp \,
  \frac{1}{ \big( R_{\rm b}^{(n)} \big)^4 } \: .
\label{eq:Blockade_n-1_red}
\end{equation}
Therefore, we get finally:
\begin{equation}
R_{\rm u, b}^{(n-1)} =
  {\Bigg( \! \frac{ 15 |C| }{4} \! \Bigg)}^{\!\! 1/4} n^{9/4} \,
  {\Bigg\{ 1 \mp \, \frac{ 15 |C| }{4} \,
  \frac{ n^9 }{ {\big( R_{\rm b}^{(n)} \big)}^{\! 4} }
  \Bigg\}}^{\! -1/4} ,
\label{eq:Position_add_zone}
\end{equation}
where the minus/plus sign refers to the points where Rydberg excitation
becomes unblocked and blocked again.

If the second term in braces in the right-hand side of
formula~(\ref{eq:Position_add_zone}) is small as compared to unity,
then this expression can be reduced to
\begin{equation}
R_{\rm u, b}^{(n-1)} \approx
  {\Bigg( \! \frac{ 15 |C| }{4} \! \Bigg)}^{\!\! 1/4} n^{9/4} \,
  {\Bigg\{ 1 \pm \, \frac{ 15 |C| }{16} \,
  \frac{ n^9 }{ {\big( R_{\rm b}^{(n)} \big)}^{\! 4} } \Bigg\}} \, .
\label{eq:Position_add_zone_red}
\end{equation}
Consequently, a center of the additional excitation zone,
corresponding to the interval $ [R_{\rm b}^{(n-1)}, R_{\rm u}^{(n-1)}] $
in figure~\ref{fig:Rydberg_blockade}, is located at the distance
\begin{equation}
R_{\rm c}^{(n-1)} =
  {\Bigg( \! \frac{ 15 |C| }{4} \! \Bigg)}^{\!\! 1/4} n^{9/4}
\label{eq:Center_add_zone}
\end{equation}
from the already excited Rydberg atom; and its characteristic width equals
\begin{equation}
\Delta R^{(n-1)} = \frac{1}{2} \,
  \Bigg( \! \frac{ 15 |C| }{4} \! \Bigg)^{\!\! 5/4}
  \frac{ n^{45/4} }{ {\big( R_{\rm b}^{(n)} \big)}^{\! 4} } \, .
\label{eq:Width_add_zone}
\end{equation}

Let us compare these expressions with the ones derived for the case of
Rydberg blockade by the uniform field~\cite{Dumin14} (they will be
designated by the subscript `unif'):
\begin{equation}
R_{\rm c, unif}^{(n-1)} =
  {\Bigg( \! \frac{ 3 C }{2} \! \Bigg)}^{\!\! 1/3} n^{7/3}
\label{eq:Center_add_zone_unif}
\end{equation}
and
\begin{equation}
\Delta R_{\rm unif}^{(n-1)} =
  {\Bigg( \! \frac{ 3 C^4 }{2} \! \Bigg)}^{\! \! 1/3}
  \frac{ n^{28/3} }{{\big( R_{\rm b}^{(n)}
  \big)}^{\! 3}} \: .
\label{eq:Width_add_zone_unif}
\end{equation}
It is quite surprising that positions of the additional excitation zone
in both cases are almost the same: the exponents of the principal
quantum number~$ n $ in formulas~(\ref{eq:Center_add_zone}) and
(\ref{eq:Center_add_zone_unif}) are very similar to each other, while
the numerical coefficients are close to unity.

Finally, it should be mentioned that the patterns of Rydberg blockade
presented in figure~\ref{fig:Rydberg_blockade} of this paper, as well as
in figure~1 of our previous article~\cite{Dumin14} involve a lot of
intersections of the energetic levels with each other.
Generally speaking, this will result in the level-anticrossing effects
and mixing between the states of different quantum numbers,
\textit{e.g.}, as discussed in~\cite{Windholz12}.
However, all such phenomena are not so important in the context of
Rydberg blockade.
The crucial item for our consideration is the presence (or absence) of
an allowed electronic state in the energy band of the exciting radiation.
The level-anticrossing effects as well as the exact identification of
the resulting energetic states are of secondary importance here.
Of course, they should be taken into consideration in a more careful
treatment of the Rydberg blockade, involving the calculation of
probabilities of the corresponding transitions.

\subsection{Comparison with the experimental data}
\label{subsec:Experimental_data}

\begin{figure}
\center{\includegraphics[width=9.5cm]{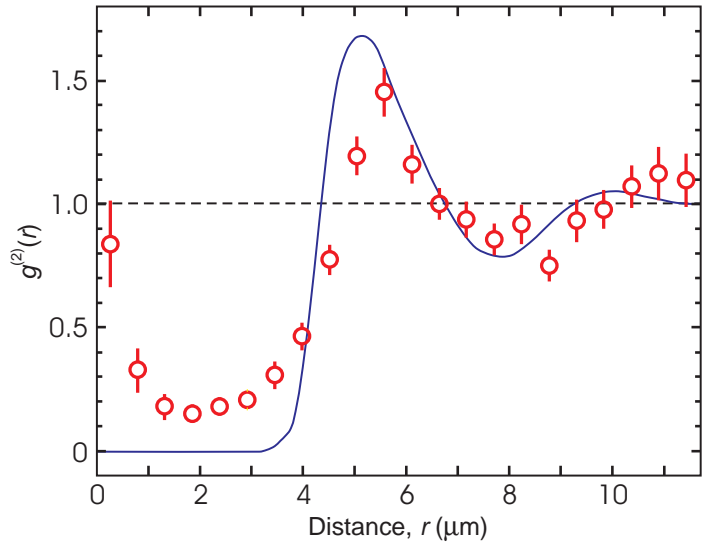}}
\caption{\label{fig:Experimental_Data}
The pair correlation function of Rydberg atoms~$ g^{(2)} (r) $
depending on the distance~$ r $ between them:
experimental data (red circles with error bars) and
the standard theoretical prediction for the Rydberg blockade
(blue solid curve).
Adapted by permission from Macmillan
Publishers Ltd: {\it Nature}, vol.~491, pp.~87--91,
\copyright 2012.}
\end{figure}

To get some numerical estimates, let us use the parameters of
experiment~\cite{Schauss12}, which seems to be the most detailed
spatially-resolved study of the Rydberg blockade available by now.
In this case, $ n = 43 $ and
$ R_{\rm b}^{(n)} \! \approx 4~{\mu}{\rm m} \approx 8{\times}10^4~{\rm a.u.} $
Then, both formulas~(\ref{eq:Center_add_zone}) and
(\ref{eq:Center_add_zone_unif}) give $ R_{\rm c}^{(n-1)} \! \approx
6{\times}10^3~{\rm a.u.} \approx 0.3~{\mu}{\rm m} $.
It is interesting that this value corresponds very well to
the position of the additional unexpected peak in the pair correlation
function of Rydberg atoms depicted in figure~3a of the above-cited paper,
which is partially reproduced in our figure~\ref{fig:Experimental_Data}.
Such a peak is evidently impossible in the ``standard'' model of Rydberg
blockade, which takes into account only the states with the same value
of the principal quantum number (the corresponding theoretical
correlation function is drawn by the solid curve).
So, this striking deviation might be associated just with formation
of the additional unblocked zone for a neighboring value of
the principal quantum number
(see interval $ [R_{\rm b}^{(n-1)}, R_{\rm u}^{(n-1)}] $ in
figure~\ref{fig:Rydberg_blockade})
rather than caused by imperfections of the measurement procedures,
as was originally suggested by the experimentalists.

It would be desirable, of course, to seek for the same peculiarity in
other experiments with Rydberg blockade.
Unfortunately, as far as we know, the other available high-precision
installations, such as~\cite{Beguin13}, cannot reach the required
sub-micron range of distances.

To avoid misunderstanding, let us emphasize that
formulas~(\ref{eq:Width_add_zone}) and (\ref{eq:Width_add_zone_unif})
can hardly be used to get the overall width of the unblocked zone,
because it is actually composed of many unblocked sublevels whose
positions are slightly shifted with respect to each other
(see figure~\ref{fig:Rydberg_blockade}).
So, this issue requires a more detailed treatment, which will be presented
elsewhere.

Finally, it should be mentioned that yet another method for treating
the Rydberg blockade in a nonuniform field might be based on
the multi-polar expansion of Hamiltonian.
The corresponding works---ranging from the detailed numerical
calculations~\cite{Madsen99,Komninos02} to the sophisticated analytical
approaches~\cite{Parzynski03}---were performed, first of all, in the context
of excitation of the high-angular-momentum Rydberg states by single
laser pulses.
Unfortunately, since this technique was not applied by now to
the problem of Rydberg blockade, we cannot make any comparison with
the results of the present paper.
Anyway, the fine spatial structure of the Rydberg blockade discussed in
our previous work~\cite{Dumin14} appears already in the approximation of
uniform inter-atomic field, and taking into account the spatial
nonuniformity results only in quantitative modifications.

\section{Conclusion}
\label{sec:Conclusion}

In summary, we derived the exact general expression for Stark splitting
of energy levels of a hydrogen-like atom by the nonuniform external
electric field.
The corresponding formula substantially refines the previously-known
result in WKB approximation.
Next, we used this expression for the multi-level treatment of the
phenomenon of Rydberg blockade and obtained the characteristic parameters
of the additional unblocked zones caused by the gradient term of
the Stark splitting.
It was found that the resulting positions of the unblocked zones are
almost the same in the different limiting cases and, unexpectedly,
coincide very well with the recent experimental measurements.

\ack

I am grateful to Profs.~J-M~Rost, R~C{\^o}t{\'e} and A~V~Gorshkov for
fruitful discussions and advises.
I am also grateful to the unknown referees for valuable bibliographic
suggestions.

\end{document}